\documentclass[12pt,preprint]{aastex}

\begin{document}

\shortauthors{Stanek}

\shorttitle{Citations vs.~length of papers}

\title{How long should an astronomical paper be to increase its {\bf
Impact}?}

\author{Krzysztof Zbigniew Stanek\altaffilmark{1}}

\altaffiltext{1}{\small Department of Astronomy, The Ohio State University, Columbus, OH 43210}

\email{kstanek@astronomy.ohio-state.edu}

\begin{abstract}

Naively, one would expect longer papers to have larger impact (i.e.,
to be cited more). I tested this expectation by selecting all ($\sim
30,000$) refereed papers from A\&A, AJ, ApJ and MNRAS published
between 2000 and 2004. These particular years were chosen so papers
analyzed would not be too ``fresh'', but at the same time length of
each article could be obtained via ADS. I find that indeed longer
papers published in these four major astronomy journals are on average
cited more, with a median number of citations increasing from 6 for
articles 2--3 pages long to about 50 for articles $\sim 50$ pages
long.  I do however observe a significant ``Letters effect'', i.e.
ApJ and A\&A articles 4 pages long are cited more than articles 5--10
pages long. Also, the very few longest ($>80$ pages) papers are
actually cited less than somewhat shorter papers.  For individual
journals, median citations per paper increase from 11 for $\sim 9,300$
A\&A papers to 14 for $\sim 5,300$ MNRAS papers, 16 for $\sim 2,550$
AJ papers, and 20 for $\sim 12,850$ ApJ papers (including ApJ Letters
and Supplement). I conclude with some semi-humorous career advice,
directed especially at first-year graduate students.

\end{abstract}

\section{Introduction}

There have been a number of publications analyzing patterns in
citations in the astronomical literature, such as ``Productivity and
impact of astronomical facilities: Three years of publications and
citation rates'' (Trimble \& Ceja 2008), ``The Importance of Being
First: Position Dependent Citation Rates on arXiv:astro-ph'' (Dietrich
2007), ``Demographic and Citation Trends in Astrophysical Journal
Papers and Preprint'' (Schwarz \& Kennicutt 2004) or ``Patterns in
Citations of Papers by American Astronomers'' (Trimble 1993), just to
mention a few.  Scanning abstracts of these publications, some career
advice emerges:

\begin{enumerate}

\item{} ``There are hot topics (cosmology, exoplanets) and not so hot
topics (binary stars, planetary nebulae)'' and ``service to the
community, in the form of catalogues and mission descriptions, is
rewarded, at least in citation numbers, if not always in other ways.''
(Trimble \& Ceja 2008)

\item{} ``We find that e-prints appearing at or near the top of the
{\tt astro-ph} mailings receive significantly more citations than
those further down the list. This difference is significant at the 7
sigma level and on average amounts to two times more citations for
papers at the top than those further down the listing.'' (Dietrich
2007)

\item{} ``Papers posted on {\tt astro-ph}  are cited more than twice as often
as those that are not posted on {\tt astro-ph}.'' (Schwarz \& Kennicutt
2004)

\item{} ``It still pays to be mature, prizewinning theorist, working
on cosmology or high-energy astrophysics at a prestigious
institution.'' (Trimble 1993)

\end{enumerate}

Some of that advice might be easier to implement than other, and
sometimes a finely timed, Wednesday afternoon 04:00:03 p.m. EST
submission to {\tt astro-ph}  produces less than desired effect (e.g., Stanek
2003, private communication).

However, to my knowledge, there has been no discussion of how long a
paper should be to maximize its impact. Is it better to write several
shorter papers or one longer paper? And serving as a referee, one of
the standard questions is ``Can this paper be made shorter?'' But
maybe it should be made longer? We have all heard a famous quote,
often attributed to Mark Twain: ``If I Had More Time I Would Write a
Shorter Letter,'' but actually from Blaise Pascal: ``Je n'ai fait
celle-ci plus longue que parce que n'ai pas eu le loisir de la faire
plus courte.'' Is not at all clear this is a valid advice when writing
astronomical papers.

To alleviate that glaring omission, I have decided to investigate the
citation vs.~paper length correlation for astronomical papers. In
Section 2 I describe the data, namely citation and page length
statistics for about 30,000 refereed astronomical papers published
between 2000 and 2004 in ApJ, A\&A, AJ and MNRAS. In Section 3 I
discuss the results, and in Section 4 I conclude with some career
advice, especially useful for first-year graduate students in
astronomy.

\addtocounter{footnote}{1}

\section{Data}

I used the ADS Abstract Service site\footnote{{\tt
http://adsabs.harvard.edu/abstract$\_$service.html}} to obtain the
page length and citations for all papers published between 2000 and
2004 (i.e., five complete years) in the four largest astronomical
journals, A\&A, AJ, ApJ and MNRAS.\footnote{These data are available
by request from the Author} I start in 2000 so that page lengths will
be reported for all papers and end with 2004 to allow time for papers
to mature (or not) and obtain citations.  The journals were chosen so
there would be plenty of papers for statistics, with the page length
being about the same for each journal.  There were 12,858 ApJ
(including ApJL and ApJS) papers, 9,275 A\&A papers, 5,330 MNRAS
papers and 2,564 AJ papers, for a total of 30,027 papers. Originally,
that number was somewhat larger, but I have removed all one-page
papers from the list, as these were mostly errata or editorials,
usually cited 0 times (although one such erratum has been cited almost
100 times). I should note that I procured these somewhat time
consuming ADS data in mid-June 2008, so individual citation numbers
might have increased a bit since then, but the overall trends should
remain the same.

\begin{figure}[p]
\plotone{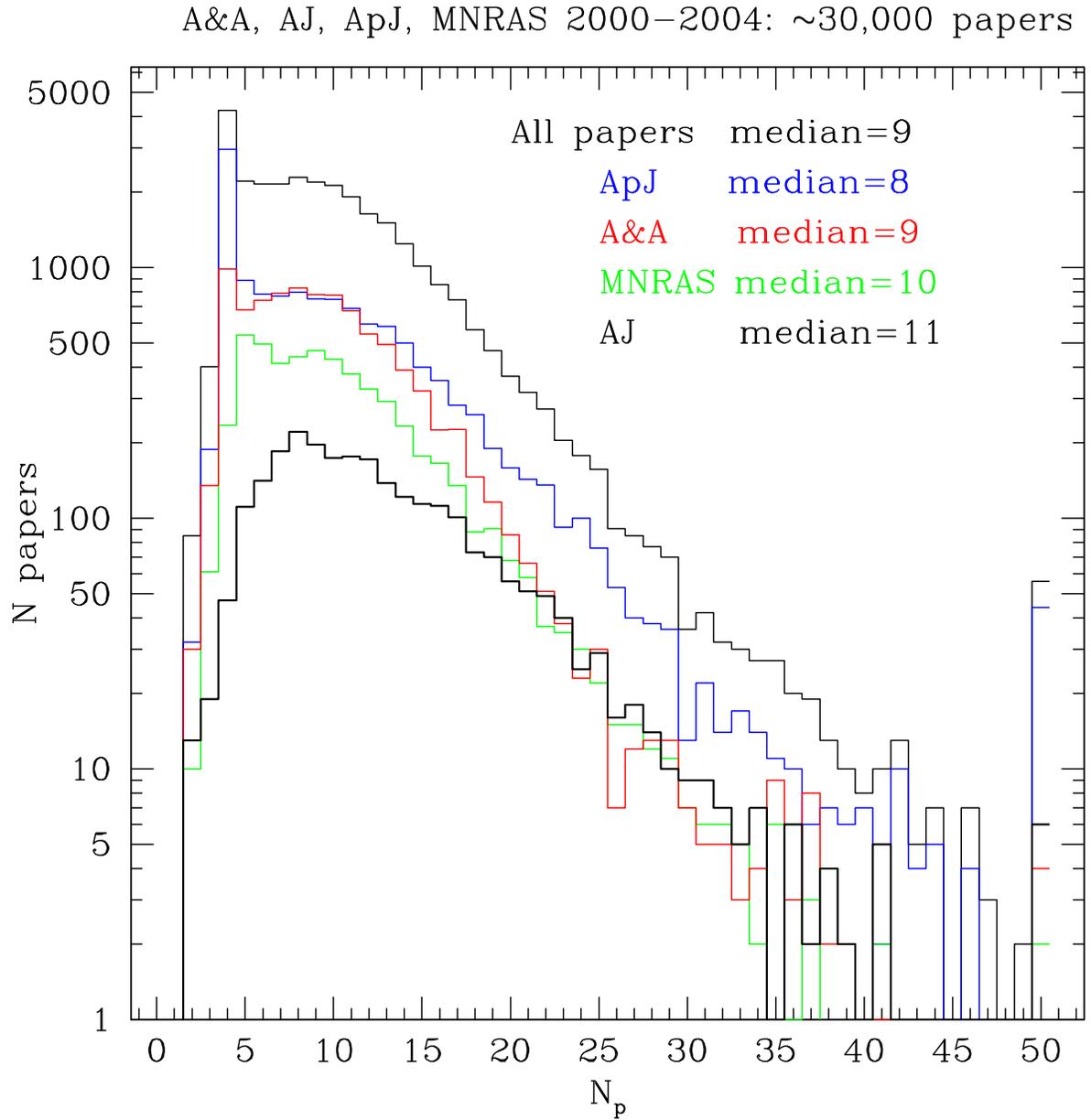}
\caption{Distribution of page length for 30,027 A\&A, AJ, ApJ, and
MNRAS papers published between 2000--2004. A significant Letters spike
can be seen at $N_p=4$.}
\label{fig1}
\end{figure}

\begin{figure}[p]
\plotone{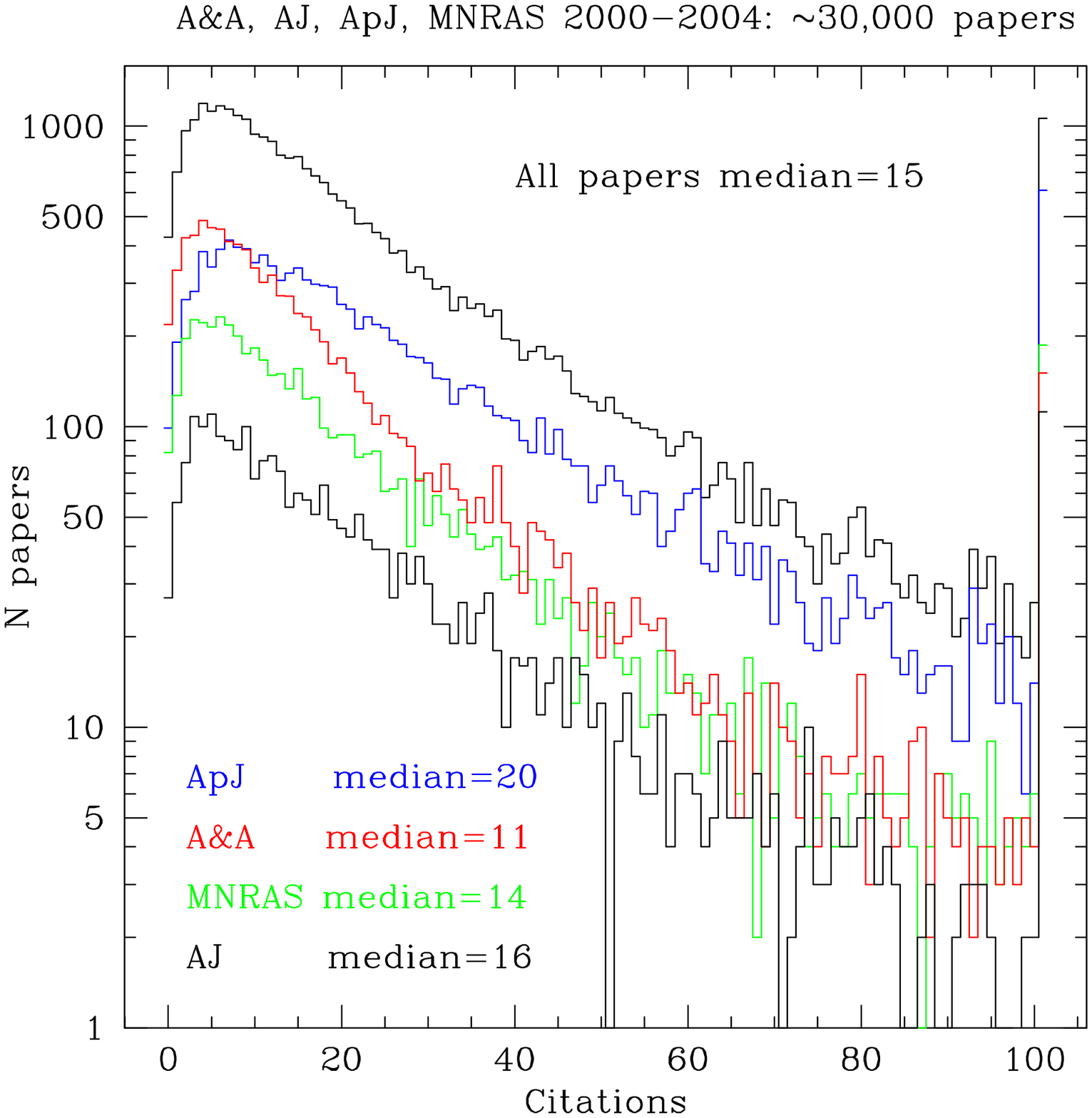}
\caption{Distribution of citations for 30,027 A\&A, AJ, ApJ and MNRAS
papers published between 2000--2004, shown for the entire sample and
for each journal separately. The median number of citations for the
entire sample is 15, ranging from 11 for A\&A to 20 for ApJ.}
\label{fig2}
\end{figure}

In Figures 1 and 2 I show the basic properties of our sample, such as
the distribution of page length (Fig.$\,$1) and distribution of
citations (Fig.$\,$2). While the distributions in presented in
Fig.$\,$1 are more or less the same, except for the expected A\&A and
ApJ Letters spike at $N_p=4$, distributions presented in Fig.$\,$2 are
not identical in shape.

While the median number of citations per paper for the entire sample
of $\sim30,000$ papers is 15, I note that $\sim3,150$ papers in the
sample have 3 or fewer citations, while $\sim1,100$ papers have 100 or
more citations, and only $\sim100$ papers have 300 or more citations.

\section{Results} 

Having both the number of journal pages and the number of citations, I
can produce a Citation-Number (of pages) Diagram (hereafter: CND),
which I present in Fig.$\,$3. For display purposes, I have added $+1$
to the number of citations for each paper (``an honorary citation''),
to avoid the $\log{citations}=-\infty$ problem
($\log{citations}=+\infty$ is not yet a problem, even for WMAP
papers).  Also, for display purposes, I add a random variable
uniformly distributed between 0 and 0.9 to both $N_p$ and number of
citations, otherwise all papers with the same number of pages and
citations would produce only one dot in the plot (since this paper is
mostly directed towards first-year graduate students, this is a very
useful trick when you have to plot data where either the data are
truly discrete or for whatever reason not enough significant digits
were preserved).

\begin{figure}[p]
\plotone{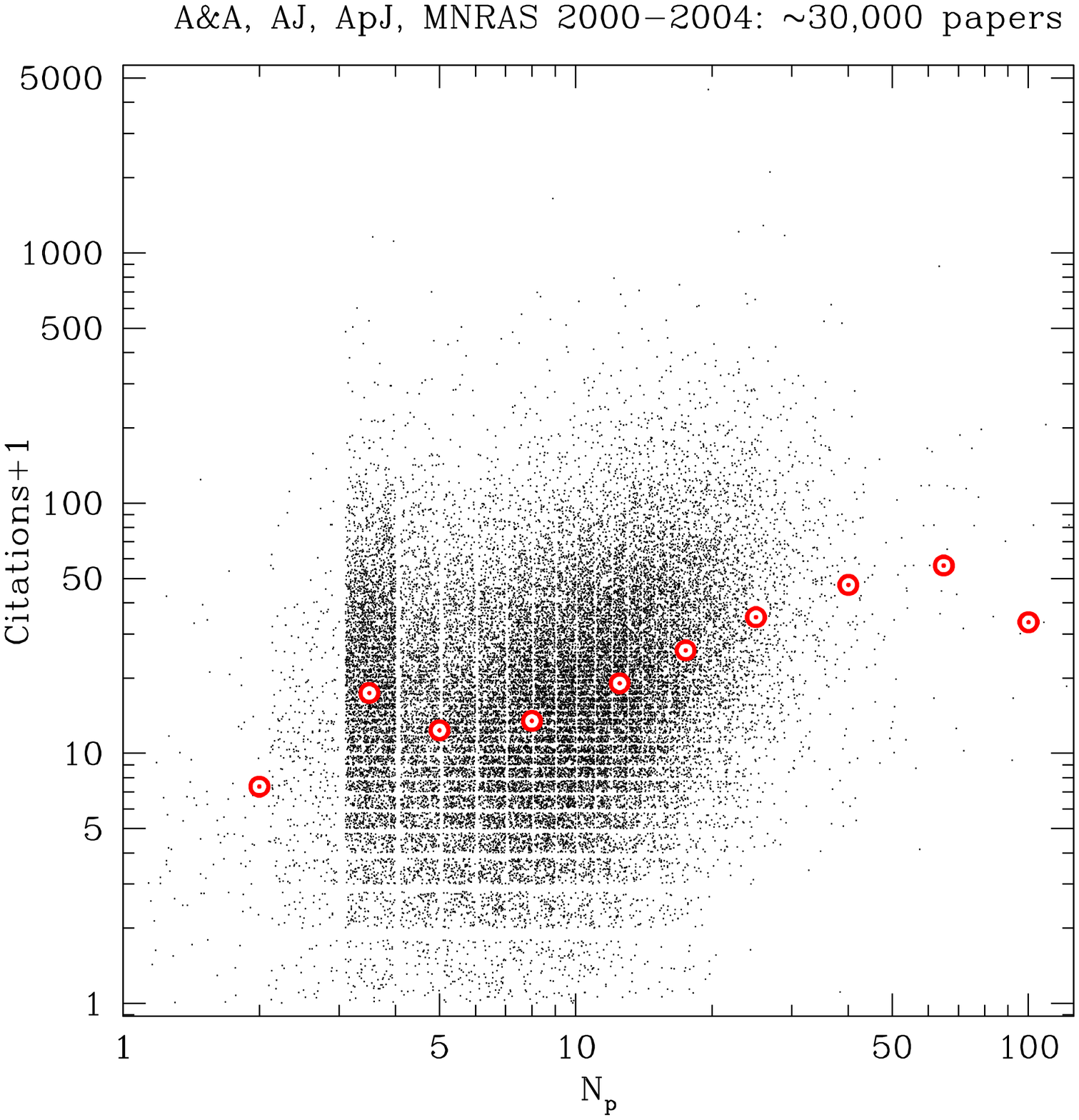}
\caption{Citations vs.~number of pages for our sample of 30,227 papers
published in A\&A, AJ, ApJ, and MNRAS between 2000--2004. For bins of
articles with a given length, the median number of citations is shown
with the large dotted circle, ranging from 6 citations for articles
2--3 pages long to $\sim50$ median citations for articles $\sim 50$
pages long.}
\label{fig3}
\end{figure}

\begin{figure}[p]
\plotone{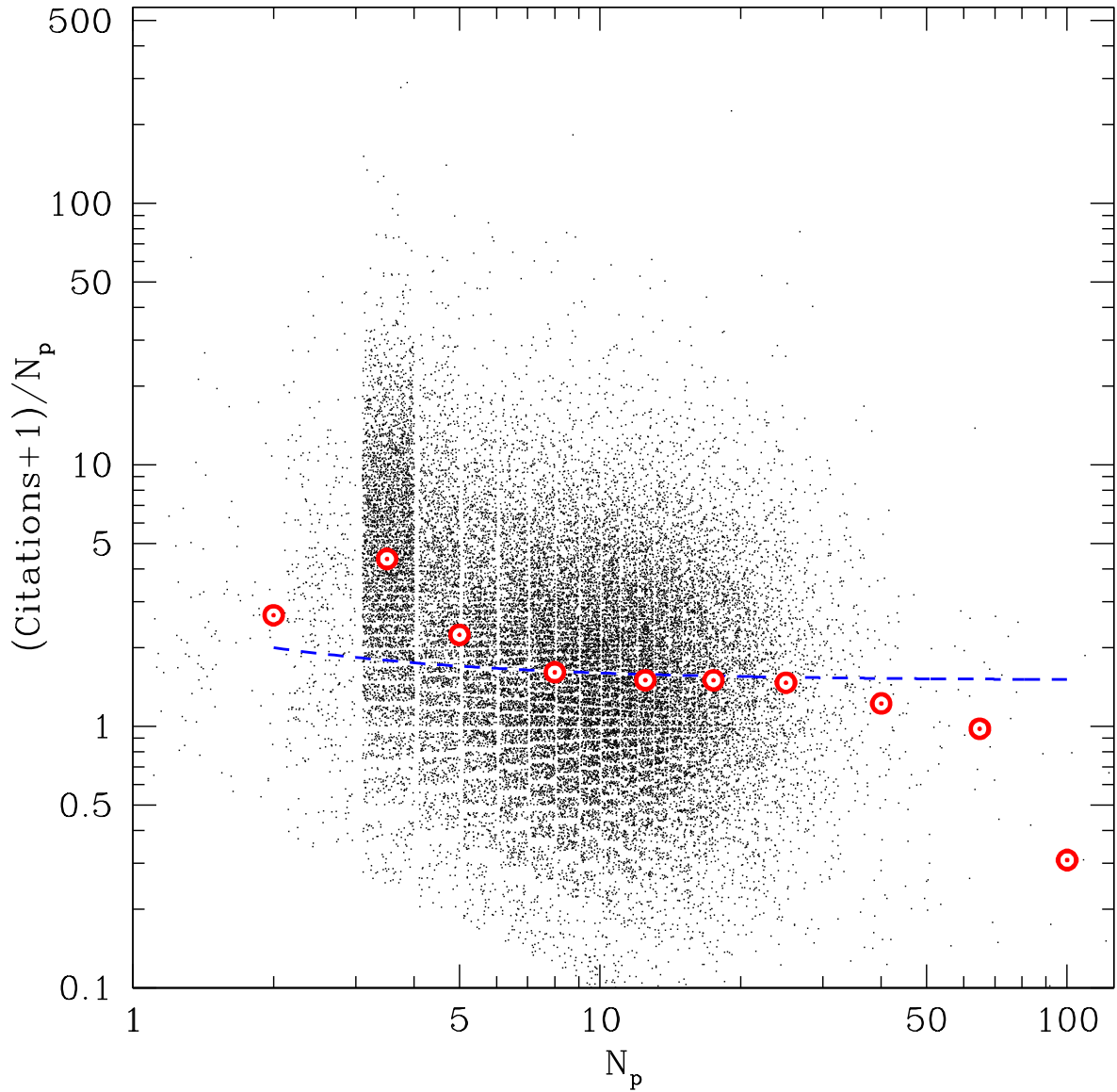}
\caption{As in Fig.$\,$3, but plotting citations/per page vs.~the
number of pages for our sample. For bins of articles with a given
length, the median number of citations per journal page is shown with
the large dotted circle. The dashed line shows a citation rate of 1.5
true citations/page, where the slope is due to the addition of one
``honorary citation'' to each paper.}
\label{fig4}
\end{figure}

The main cloud of points in the resulting CND (Fig.$\,$3) looks like
the state of Ohio (USA),\footnote{\tt
http://www.smart-traveler.info/sitebuildercontent/sitebuilderpictures/map$\_$of$\_$ohio.gif}
with very short papers with some citations located in Indiana, papers
with none or very few citations located in Kentucky (short) and West
Virginia (long), papers with very many citations located in Michigan
(short), Lake Erie and Canada, and very long papers located in
Pennsylvania.  Readers can amuse themselves by locating their various
articles on the CND and mentioning casually at their Coffee ``I have
three papers in Canada'', ``my paper just fell into the Lake Erie'' or
``my paper has finally moved from Coshocton County to Tuscarawas
County.''

For bins of journals articles with a given length, the median number
of citations is shown by large open circles. These medians range from
6 for articles 2--3 pages long to about 50 for articles $\sim 50$
pages long. There is a local maximum corresponding to papers 4 pages
long, which I will show below is due to A\&A and ApJ Letters.  Also,
the very few longest ($>80$ pages) papers are actually cited less than
somewhat shorter papers, although there are few such very long
papers. One obvious thing to notice is that, as expected, for any
given length of the paper there is a very wide range in the number of
citations.

Impact per journal page (Fig.$\,$4) is fairly flat at $\sim$1.5
citations/page for papers longer than 6 pages and shorter than 50
pages, but it peaks at $\sim$4 citations/page for papers 4 pages long
and drops rather steeply for the longest papers.  In the entire sample
of $\sim 30,000$ papers there are only 10 papers exceeding the rate of
100 citations/page and 54 papers exceeding the rate of 50
citations/page.

Seeing the morphology of the overall CND, I then decided to produce
individual CNDs for each of the four journals, shown in Fig.$\,$5. ApJ
data are shown in the upper-right corner of Fig.$\,$5, and the dashed
line corresponding to bin medians for ApJ is then also shown in the
panels for the three other journals. Both ApJ and A\&A show a local
maximum at $N_p=4$, most likely due to Letters, while no such local
maximum is present for AJ or MNRAS papers. 

\begin{figure}[p]
\plotone{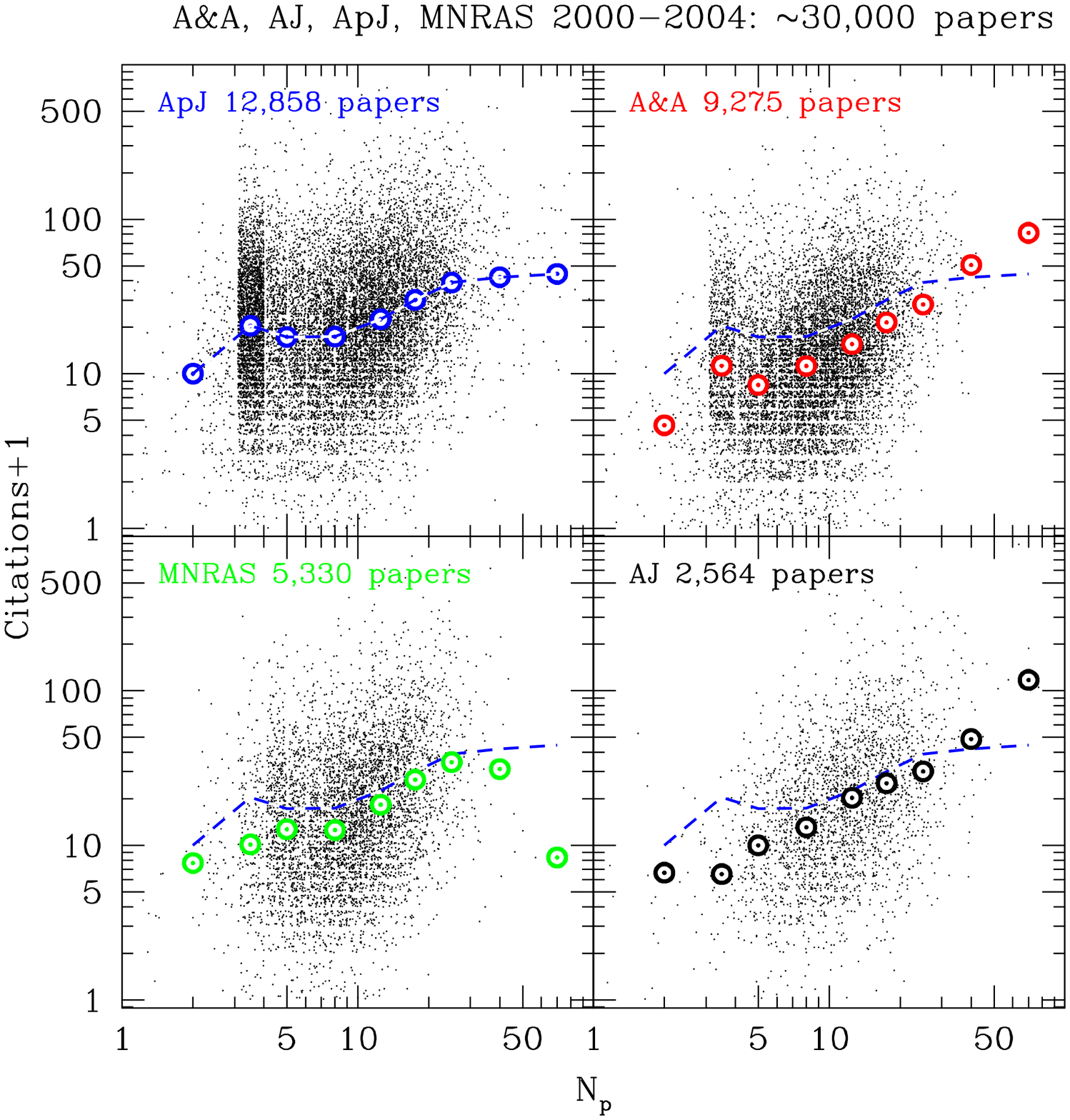}
\caption{As in Fig.$\,$3, but now shown separately for each journal.
Dashed line corresponding to medians in each ApJ length bin is shown
in each panel for comparison with the other journals.}
\label{fig5}
\end{figure}

\section{Semi-humorous Career Advice}

For non-humorous career advice in astronomy, see, e.g., ``So you want
to be a professional astronomer!'' (Forbes 2008) or ``The Production
Rate and Employment of Ph.D. Astronomers'' (Metcalfe 2008).  Here is
my semi-humorous publishing advice, directed especially towards
first-year graduate students\footnote{There is no guarantee that any
of this advice will be helpful in your career; it is provided as a
public service and a source of possible amusement. No warranty is
expressed or implied.  The Ohio State University disavows any
responsibility for the information contained herein.} (this advice
results from the current study combined with conclusions presented in
other studies cited here):

\begin{itemize} 

\item{} Initially, write and post to {\tt astro-ph} as many
first-author papers as possible and do not yet worry (too much) about
their length or citations. As philosophers have asked since the time
of ancient Greeks and maybe even before, ``If you did the research and
did not post it on astro-ph, did it really happen?''  At this point in
your career first author papers are crucially important for you, due
to the prevailing system of attributing credit in astronomy and the
way papers are referred to using the name of the first author (name
recognition).  Make sure you submit your papers to {\tt astro-ph}
archive just after 4 p.m.~US Eastern time\footnote{\tt
http://arxiv.org/localtime} on Wednesday (and avoid, if you can,
posting your paper such that it appears on Tuesday)---this way your
papers will on average be cited about 3 times more than papers only
published in journals (Dietrich 2007; Schwarz \& Kennicutt 2004).

\item{} Also early in your career, it makes sense to write and post on
astro-ph your conference contributions, as they increase your name
recognition, but longer term, conference proceedings papers are cited
20 times less often than the average ApJ paper (Schwarz \& Kennicutt
2004), and in fact median number of citations to a conference
contribution is either 0 or 1.

\item{} Later in your career you will worry more about your total
number of citations, but you will also discover that you somehow need
to find money to pay for the journal page charges. This is when you
want to increase the fraction of Letters in your publication
portfolio, which increases the number of citations per unit of
currency.\footnote{Monthly Notices of the Royal Astronomical Society
has no page charges, except for colour in the printed version. A\&A
has page charges at the rate of 100 euros per printed page, but
charges are not requested if the first author is affiliated with one
of the countries that sponsor A\&A. AJ and ApJ charge between \$105
and \$195 (paper submission) per page, or you can just send them 20
euros in cash.}

\item{} At a still later point in your career you will find out about
the $h$-index (Hirsch 2005) and you will become obsessed with it,
possibly listing its value on your CV and/or website and telling other
astronomers that your $h$ is bigger than their $h$. You will therefore
want to start writing longer papers, as these will be more likely to
have more citations overall, increasing your value of $h$-index. At
this point of your career you do not have to worry as much as before
about being the first author on your papers, but you will probably
have to find money to pay for the page charges for your first-author
junior collaborators, who face a different optimization problem.

\end{itemize}

\acknowledgments

I thank the late Professor Bohdan Paczy\'nski for his valuable advice
and insight over the years, including always valid ``Many astronomers
confuse difficult with interesting'', which partially inspired this
paper.  I would like to thank the participants of the morning
``Astronomy Coffee'' at the Department of Astronomy, The Ohio State
University, for the daily and lively {\tt astro-ph} discussion, one of
which finally prompted me to produce this posting. However, they do
not deserve any blame in the unlikely case you lack a sense of
humor. I thank Chris Kochanek and Molly Peeples for their useful
comments on an earlier version of this paper.  The data and plots
presented in this paper are real, despite the (mostly intended)
humorous style of writing. This paper will not be submitted to any
journal, but please feel free to cite it as often as possible, or
better yet cite my regular astronomical papers.

\end{document}